# Relationship between the magnetic hyperfine field and the magnetic moment

S. M. Dubiel*

Faculty of Physics and Applied Computer Science, AGH University of Science and Technology, al. A. Mickiewicza 30, PL-30-440 Kraków, Poland

**Abstract**

Based on experimental data it is shown, for some chosen alloys and compounds of iron, that there is no one unique relationship between the $^{57}$Fe-site magnetic hyperfine field, $B_{hf}$, and the magnetic moment per Fe atom, $\mu$. Instead, the $B_{hf}$-$\mu$ plot consists of several branches, each of them being characteristic of a given alloy or compound. Consequently, the effective proportionality constant (hyperfine coupling constant) depends on the alloy system or compound, and for a given alloy system or compound it depends on the composition or even on the lattice site. Consequently, the scaling of $B_{hf}$ into the underlying $\mu$ cannot be done *a priopri*.

- dubiel@novell.ftj.agh.edu.pl

## 1. Introduction

Magnetic alloy systems and compounds of iron are often investigated with the use of Mössbauer spectroscopy and the magnetic hyperfine field, $B_{hf}$ being the main spectral parameter. A question arises whether or not an information on the underlying magnetic moment, $\mu$, can be derived from $B_{hf}$. A rather quite frequent practice, applied to recalculate $B_{hf}$ into $\mu$, has been to simply divide the measured value of $B_{hf}$ by ~15 T/$\mu_B$, a figure obtained by dividing the value of $B_{hf}$ measured for a metallic Fe (33.9 T at 4 K), by the value of the magnetic moment per Fe atom in iron (2.2 $\mu_B$). This procedure was applied for various alloy systems and compounds having not only different compositions but also different crystallographic structures [1-7], despite theoretical calculations carried out as early as in 1961 clearly demonstrated that even for the pure metallic iron the two quantities are not proportional to each other [8]. It is also known from measurements that the temperature dependence of $B_{hf}$ do not, in general, follow that of the magnetization, the difference being temperature dependant [9-11]. This means that the proportionality constant between the two quantities is also temperature dependent. As will be shown below on several examples of Fe-containing alloys and compounds, the effective hyperfine coupling constant, $A$, i.e. the figure obtained by dividing the measured $^{57}$Fe-site hyperfine field by the magnetic moment per Fe atom determined either from magnetization or neutron measurements depends on the alloy system, and for a given alloy it depends on its composition and degree of order. For intermetallic compounds, the constant depends also on a crystallographic site. An extreme example seems to be a $Ru_xFe_ySi$ system for which the effective coupling constant changes between 189.3 T/$\mu_B$ for $RuFeSi_{0.5}$ and 27.7 T/$\mu_B$ for RuFeSi [1]. In other words, there is no one definite value of this constant, hence the relationship between $B_{hf}$ and $\mu$ is not universal which means that the rescaling of $B_{hf}$ into $\mu$ cannot be done *a priori*.

## 2. Theoretical background

Magnetic hyperfine field as measured, for example by Mössbauer spectroscopy, $B_{hf}$, can be expressed as a vectorial summ of three components:

$$\vec{B}_{hf} = \vec{B}_d + \vec{B}_o + \vec{B}_c \qquad (1)$$

where $B_d$ is a dipole term, $B_o$ is an orbital term and $B_c$ is the Fermi contact term. For iron alloy systems and compounds, the orbital term is small because the expectation value of the orbital momentum, $<L>$, is quenched by the crystalline field [12], and the dipolar term is of the order of 2 T [13] i.e. the two terms are much smaller than $B_{hf}$ (equal to 33 T for metallic Fe at room temperature). In these circumstances with a good approximation $B_{hf} = B_c$.

$B_c$ has its origin in a different density of s-like electrons with spin-up (↑) and spin-down (↓) within the volume of nucleus, $\rho(0)$. Thus neglecting the first two terms in equ. (1), $B_{hf}$ can be expressed as follows:

$$B_{hf} = a' \sum_i \left[ \rho_i^{\uparrow}(0) - \rho_i^{\downarrow}(0) \right] \qquad (2)$$

where $a'$ is a proportionality constant. The $B_c$ term is often regarded as consisting of two contributions:

$$B_c = B_{cp} + B_{cep} \tag{3}$$

with $B_{cp}$ representing the field due to a polarization of core (1s, 2s, 3s) electrons and $B_{cep}$ representing the one due to the polarization of band (conduction or valence) electrons (4s, 3d, 4p). The reason for such separation follows from theoretical calculations by Watson and Freeman according to which only the first term is proportional to the 3d shell magnetic moment, $\mu_d$ [8]. Having this in mind, $B_{hf}$ can be expressed as follows:

$$B_c = a\mu_d + b\left[\rho^{\uparrow}(0) - \rho^{\downarrow}(0)\right] \tag{4}$$

In equ. (4) $\rho(0)$ stands for the total band (conduction or valence) electron contact density. $B_c$ - $\mu_d$ linearity was also confirmed theoretically by other authors [14-17]. However, the value of the proportionality constant, $a$, depends on theoretical approach. In this respect the most detailed calculations were carried out by Lindgren and Sjøstrøm who calculated $B_{cp}$ (1s, 2s, 3s) and $B_{cep}$ (4s, 3d, 4p) terms for five different exchange correlation potentials both for a band iron and a free Fe atom ($3d^7 4s^1$), using relativistic and non-relativistic approximation [6]. They found that $a$ was between $-17.1$ T/$\mu_B$ and $-12.1$ T/$\mu_B$ for the relativistic band calculations, and between $-12.1$ T/$\mu_B$ and $-9.3$ T/$\mu_B$ for non-relativistic free atom calculations, depending on the exchange potential. The corresponding constant for the total field i.e. $B_c$ was between $-20.25$ T/$\mu_B$ and $-15.0$ T/$\mu_B$. The proportionality constant $b$ has yet different value; e. g. according to a recent paper, in which the full-potential linearized augmented plane wave method based on the density function theory was used, it is equal to 9 T/$\mu_B$ [17], while according to others, $b = 26.1$ T/$\mu_B$ [15]. Calculations performed with the spin-polarized discrete variational method for clusters of 15 atoms in Fe-Cr alloys showed that both the value and the sign of the proportionality constant strongly depend on the atomic configurations, and it ranges between 12.0 T/$\mu_B$ and $-12.5$ T/$\mu_B$ depending on the number of Fe and Cr atoms in the first two coordination shells [18]. Neglecting these differences among various approaches applied in the theoretical calculation it is clear that the proportionality between the measured total hyperfine field, $B_{hf}$, and the magnetic moment, $\mu$, often used in an interpretation of Mössbauer data should be used with caution since the band (conduction or valence) term is not necessarily proportional to the magnetic moment. The deviation from the proportionality should be greater for systems where the $B_{cep}$ term is relatively large i. e. in itinerant magnetic systems. Conversely, for magnetic systems with a well localized magnetic moment, the relationship should be rather linear. In fact, the latter seems to be the case for R-Fe compounds [19]. In particular, for various Y-Fe ones for which the value of the proportionality constant between the Fe-site hyperfine field and the Fe-site magnetic moment is ~15 T/$\mu_B$ [20]. On the other hand, the are also exceptions in this group of compounds, like $YFe_6Sn_6$, for example, where the $B_{hf}$ - $\mu$ relationship shows a pronounced curvature which likely originates from the fact that only the core electron contribution is proportional to the Fe magnetic moment while the conduction electron term is not [19].

## 2. Experimental relationship

### 2. 1. Hyperfine field versus magnetic moment

A relationship between $B_{hf}$ and $\mu$ as measured on three types of iron alloy and compound systems viz. (a) disordered binary Fe-M alloys with M = Co, Cr and Si, (b) ordered ($DO_3$) $Fe_3Si$ and $Fe_3Al$ intermetallic compounds with Fe atoms substitute by Co, Cr, Ni, Mn and Ru and (c) sigma-phase FeCr non-stoichiometric compound is presented in Fig. 1. The plot was constructed based on the data published in [21,22,29,30-35] It is clear that the data do not follow any unique linear relationship. Instead, different branches, similar to the Slater-Pauling curves known for magnetic moments or Curie temperatures, can be observed. The types (a) and (b) of alloys have their own branches and within type (a) each of the three investigated alloys has its own branch. On the other hand, the data for the sigma-Fe-Cr lie well on the branch characteristic of the disordered Fe-Si alloys. This branch and the one for the disordered Fe-Cr alloys have a quasi-linear character, though their slope is significantly different. However, the branch for the disordered Fe-Co alloys has a pronounced curvature.

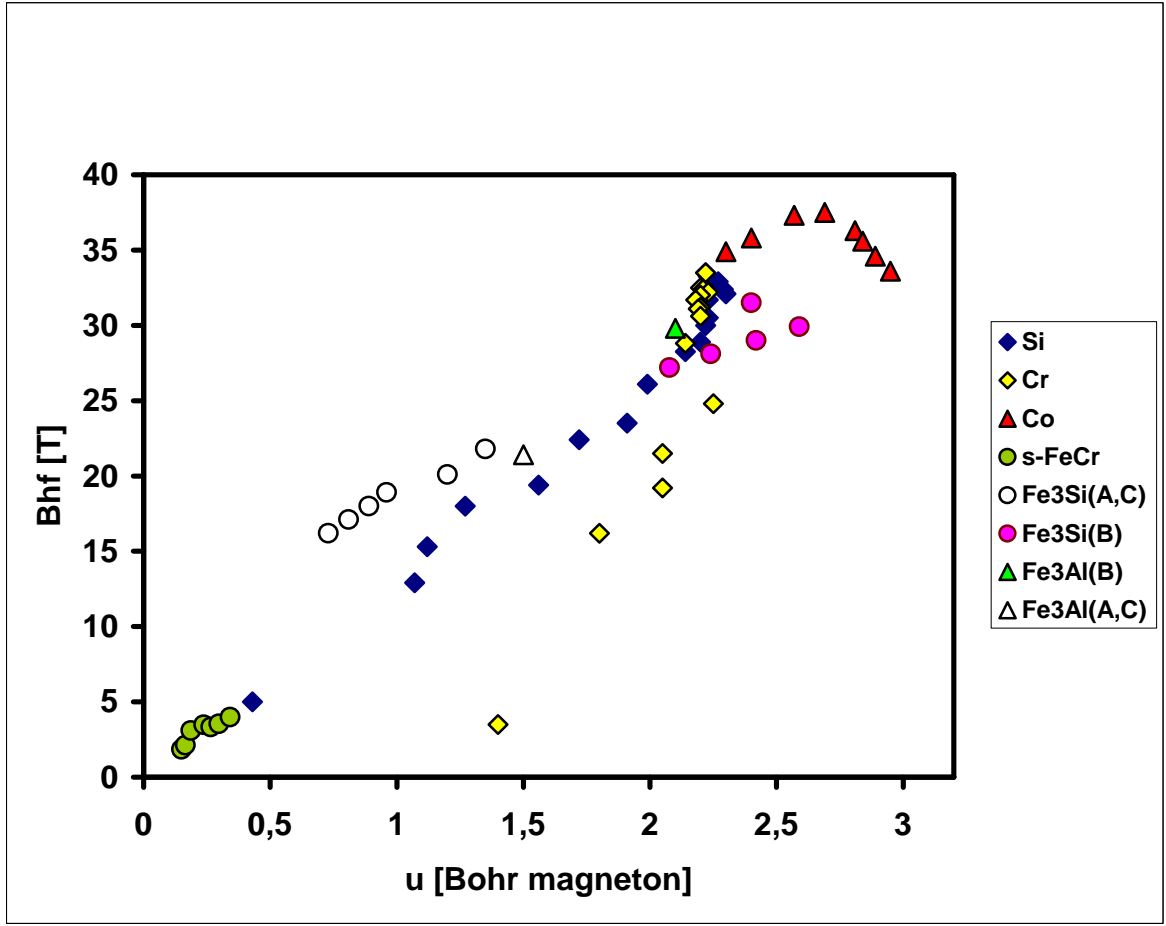

Fig. 1 Hyperfine field, $B_{hf}$, versus magnetic moment per Fe atom, $\mu$, for (a) disordered Fe-M alloys (M = Co, Cr, Si), (b) ordered ($DO_3$) $Fe_{3-x}X_xSi$ (X = Co) and $Fe_3Al$ alloy and (c) sigma-phase Fe-Cr alloys.

To show further that for a given class of alloys the $B_{hf}$ - $\mu$ relation, hence the value of the effective coupling constant, *A*, depends on the composition, such relation is illustrated in Fig. 2 only for the sigma-phase in Fe-Cr and Fe-V systems.

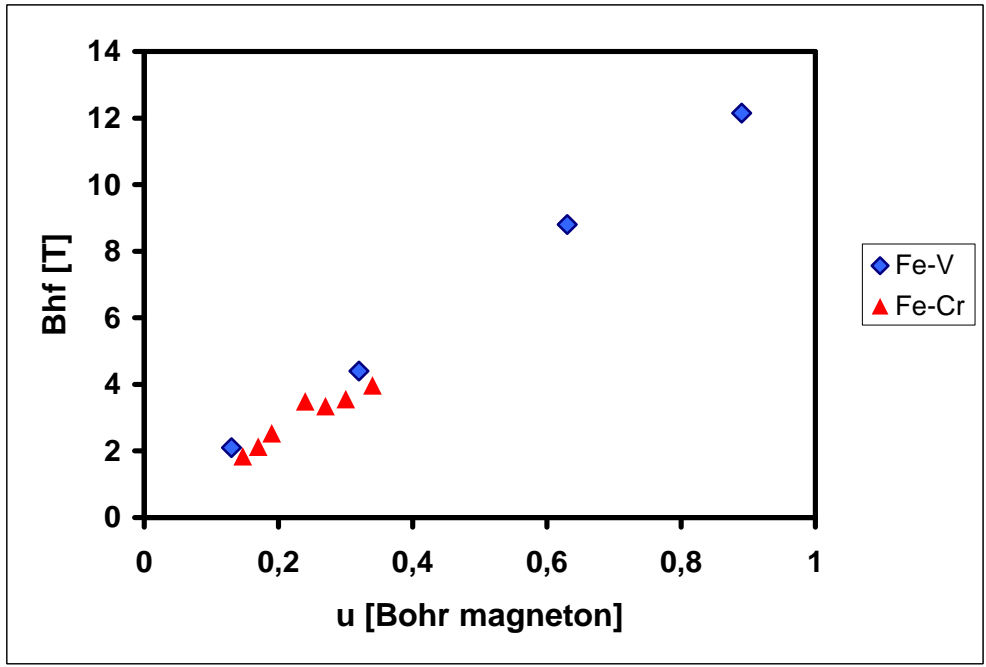

Fig. 2 Average hyperfine field, $B_{hf}$, versus average magnetic moment per Fe site, $\mu$, for the sigma-phase in Fe-Cr and Fe-V systems [31].

Here the $B_{hf}$ - $\mu$ relation is non-linear for the Fe-Cr system and perfectly linear over a much wider range of composition for the Fe-V system.

Further discussion of the issue will depict the value of the effective proportionality constant (hyperfine coupling constant), $A$, between the measured hyperfine field and the magnetic moment per Fe site (for disordered alloys and for the sigma-phase both quantities are average ones) i. e. calculated assuming $B_{hf} = A \mu$.

2. **2. Disordered binary alloys**

Figure 3 illustrates $A$ as a function of $x$ for $Fe_{100-x}M_x$ (M=Co, Cr and Si) obtained from the data published in [21,22,29,30-35]. It is obvious that $A$ is characteristic of a given alloy and it is also concentration dependent i.e. it more or less linearly decreases with $x$. For $x \leq \sim 18$, $A_{FeCr} \approx A_{FeSi}$ but $A_{FeCo} > A_{FeCr,FeSi}$. For $x \geq \sim 75$, $A_{FeCr}$ shows a steeper decrease with $x$ and its value at $x \approx 95$ is about 7-8 times smaller than that for a pure iron. For very diluted $Fe_{100-x}Cr_x$ alloys ($x < 1$), Fe atoms posses the magnetic moment of ~1.4 $\mu_{B+,}$ yet the hyperfine field is as small as 3.5 T [33], consequently the effective hyperfine coupling constant is equal only to 1.5 T/$\mu_B$ i.e. ten times less than for $x = 0$.

**2. 3. Ordered alloys and compounds**

Archetypal examples of the $DO_3$ superstructue are $Fe_3Al$ and $Fe_3Si$ compounds. Here the effective proportionality constant, $A$, that was calculated for the former based on the results presented in [30,31], and for the latter on the results published in [38], is equal to 16.6 T/$\mu_B$ in $Fe_3Al$, and 15.4 T/$\mu_B$ in $Fe_3Si$ for sites A and C against 14.7 T/$\mu_B$ in $Fe_3Al$ and 16.1 T/$\mu_B$ in $Fe_3Si$ for site B. Yet a greater difference between the sites for the $Fe_3Si$ compound follows from the results given in [9]. Here $A_{A,C} = 17.0$ T/$\mu_B$ against $A_B = 13.3$ T/$\mu_B$. This illustrate well the fact that $A$ in a given compound or ordered alloy can be even site dependent. On the other hand, the kind of element with which Fe makes the compound also seems to be important as far as the value of $A$ is concerned. In particular, for $Fe_3Ge$ the site average value of $A = 12.3$ T/$\mu_B$ as determined from the data published in [39,40] which is significantly less than the value found for $Fe_3Al$ and $Fe_3Si$. To further investigate the issue we consider the $Fe_3Si$ compound in which Fe atoms have been substituted by Co atoms.

$A$ versus $x$ for $Fe_{3-x}Co_xSi$ compounds obtained from the data published in [29], is presented in Fig. 4. The most striking feature to be noticed here is a clear dependence of $A$ on the crystallographic site, namely $A_B > A_{A,C}$. Other interesting feature that can easily be noticed is a different concentration dependence of $A_B$ and $A_{A,C}$. The former increases with $x$, reaches its maximum at $x \approx 0.8$ and falls again for greater $x$. The latter initially decreases with $x$, reaches its minimum at $x \approx 0.5$ to increase slowly for larger $x$ – values. In any case, for all $x$-values, except $x = 0$, the values of $A$ differ significantly for the two sites and the maximum difference equals to 11.5 T/$\mu_B$ for $x \approx 0.8 – 1$ i.e at this composition $A_B$ is twice as big as $A_{A,C}$.

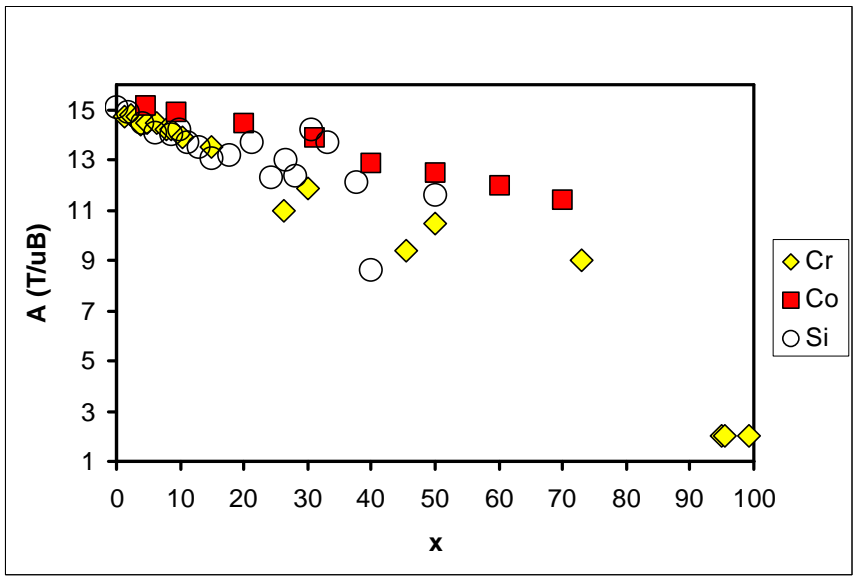

Fig. 3 Effective hyperfine coupling constant, $A$, for disordered $Fe_{100-x}M_x$ alloys with M = Co, Cr and Si versus $x$, as determined based on the data published in the literature.

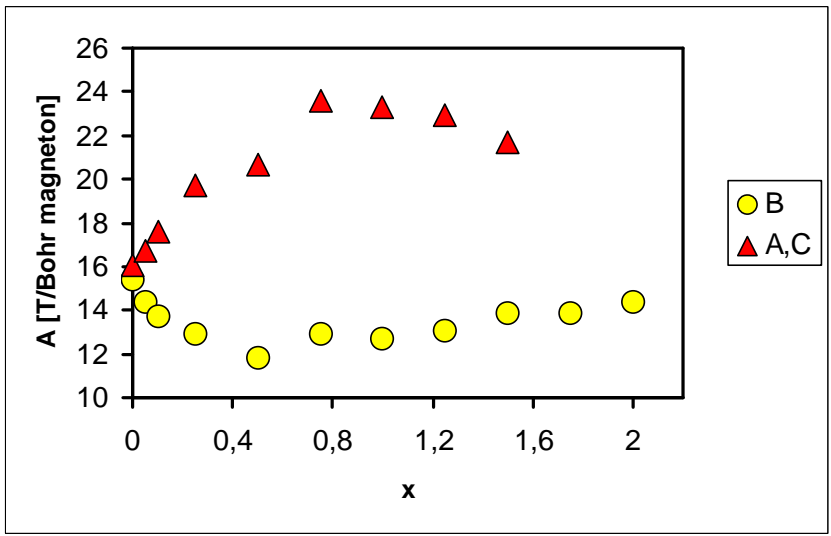

Fig. 4. Effective hyperfine coupling constant, $A$, for ordered (DO$_3$) Fe$_{3-x}$Co$_x$Si alloys versus $x$, as deduced from the data presented in [29].

Quite similar behaviour, as illustrated in Fig. 5, exhibits the Fe$_3$Si alloy with Cr and Ni atoms substituted for Fe. On the other hand, the $A_{A,C}$ – values for M = Mn do not show any systematic character. This also contrasts with M = Al where, as it follows from [41], $A_{A,C} \approx A_B \approx 18$ T/$\mu_B$.

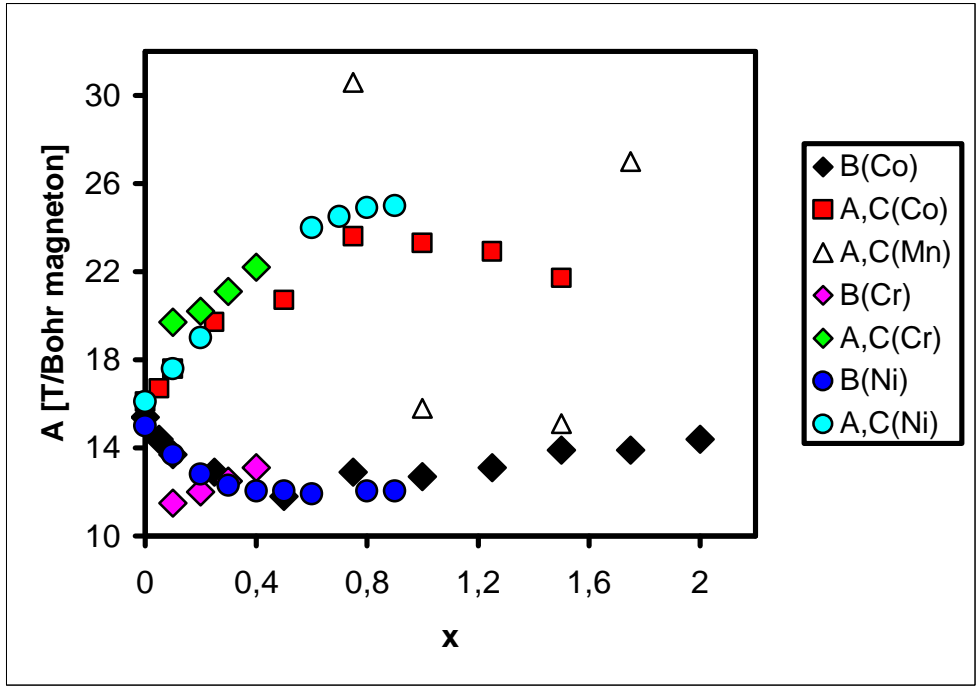

Fig. 5. Effective hyperfine coupling constant, $A$, for ordered (DO$_3$) Fe$_{3-x}$T$_x$Si alloys (T=Co, Cr, Ni, Mn) versus $x$.

In the case of Fe$_{3-x}$Ru$_x$Si the magnetic moment was determined from the magnetization measurements, so only its average value per Fe atom is known [23]. Consequently, no distinction between the two sites could have been made. The average value of $A$ over the two sites versus $x$ is plotted in Fig. 6, from which it is clear that $A \approx 14$-$15$ T/$\mu_B$ up to $x \approx 1.4$, followed by a steep quasi-linear increase. The difference between the minimum and the maximum value of $A$ is here one order of magnitude.

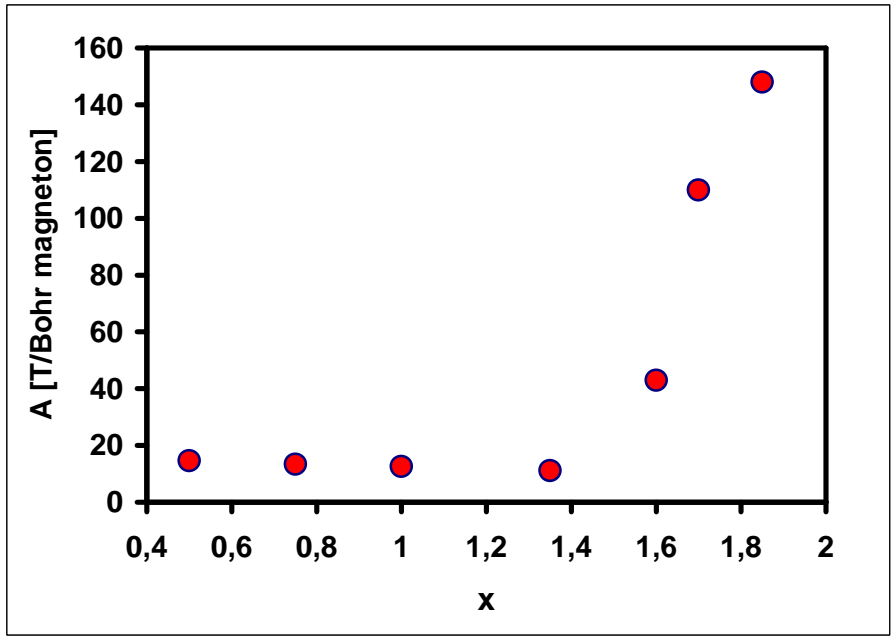

Fig. 6. Effective hyperfine coupling constant, *A*, for ordered (DO$_3$) Fe$_{3-x}$Ru$_x$Si alloys versus *x*.

Alike effect can be seen in Fig. 7 for DO$_3$ ordered Fe$_{100-x}$Al$_x$ alloys with $22.5 \leq x \leq 34$ as derived from the data presented elsewhere [24]. One can readily see that also here *A* is concentration dependent, and it increases with *x* from 14 T/$\mu_B$ at $x \approx 23$ to 45 T/$\mu_B$ at $x \approx 34$.

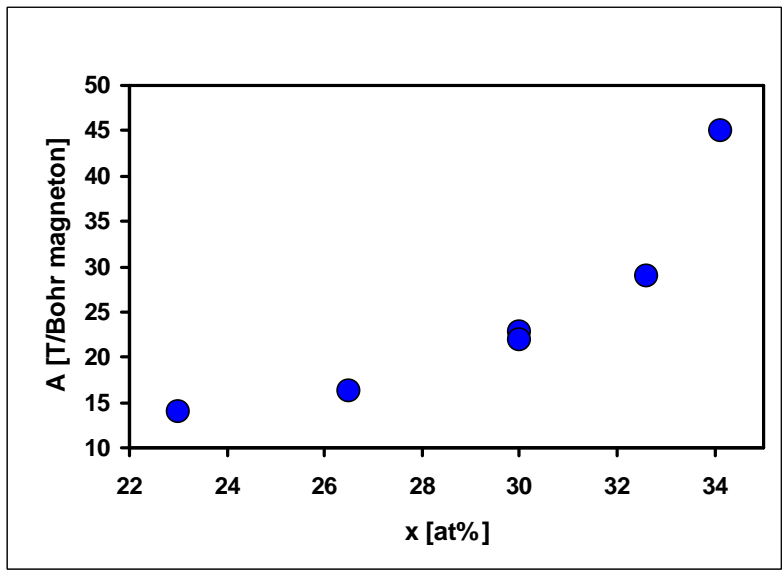

Fig. 7 Effective proportionality constant, *A*, between $B_{hf}$ and $\mu$ for DO$_3$ ordered Fe$_{100-x}$Al$_x$ versus *x*.

Similar effect of the composition on the effective hyperfine coupling constant is shown in B2 ordered alloys. Figure 8 illustrates this behavior for disordered and ordered Fe$_{100-x}$Co$_x$ alloys.

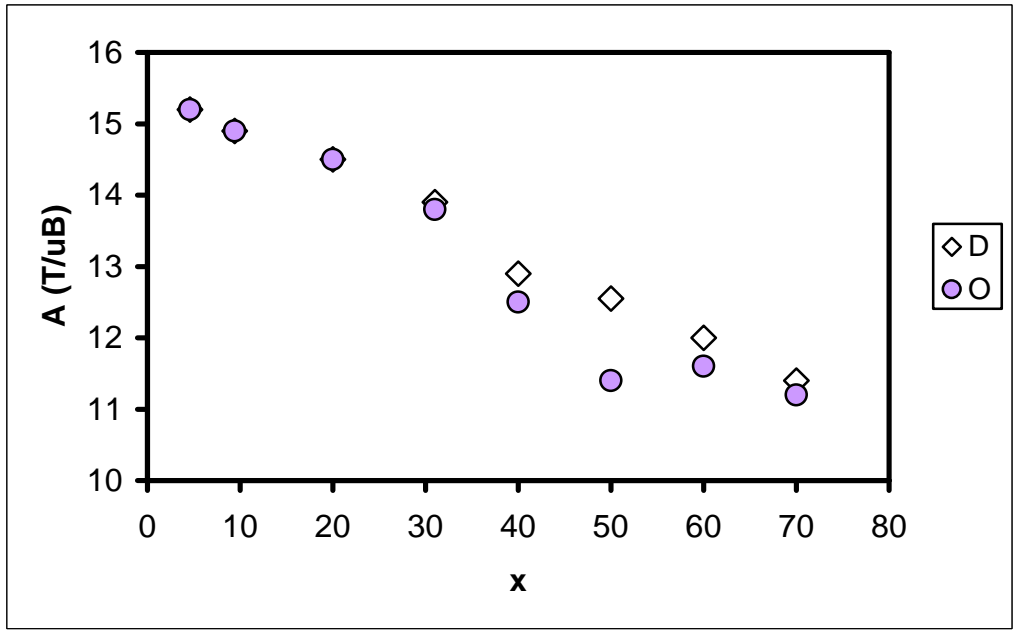

Fig. 8 Effective hyperfine coupling constant, *A,* versus *x* for disordered (D) and ordered (O) $Fe_{100-x}Co_x$ alloys. The greatest difference in *A* can be seen around *x* = 50 i.e. at the composition where the degree of the B2 order is at maximum.

Much smaller *A* - values can be found in other ordered alloys or compounds of iron. In particular, using the appropriate data for $Ni_3Fe$ from [42,43] one arrives at *A* = 9.1 T/$\mu_B$, and at 11.1 T/$\mu_B$ or 9.4 T/$\mu_B$ for site 1 and 2, respectively, in $Fe_2P$ [44], and finally at 9.5 T/$\mu_B$ for $Fe_3Sn_2$ [45].

### 3. 3. Sigma-phase Fe-Cr and Fe-V alloys

Sigma-phase alloys have a very complex crystallographic structure with 30 atoms distributed over five different lattice sites. Consequently, only the average values both of the hyperfine field, $B_{hf}$, as well as those of the magnetic moment per Fe atoms, $\mu$, could have been determined. The average value of the hyperfine coupling constant, *A*, derived from this data for the sigma-phase in the Fe-Cr system is plotted in Fig. 9. As can be seen it is concentration dependent with a maximum of ~ 14.5 T/$\mu_B$ at x ≈ 47 and a minimum of ~ 11.5 T/$\mu_B$ at *x* ≈ 44.5. On the other hand, the value of *A* derived in a similar way for the sigma in the Fe-V system is constant within V concentration of 34 – 60, and equal to 13.15 T/$\mu_B$.

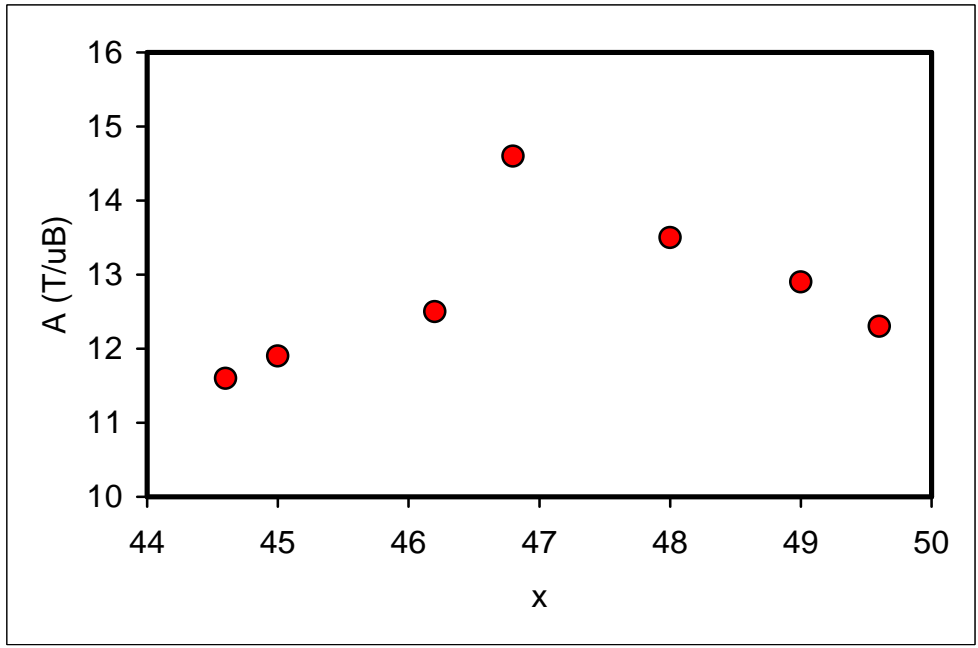

Fig. 9 Average hyperfine coupling constant, *A*, versus Cr concentration, *x*, in the sigma-$Fe_{100-x}Cr_x$ alloys [22].

### 4. 4. Amorphous alloys

The relationship between $B_{hf}$ and $\mu$ was also investigated in various amorphous alloys of iron. Here in many instances a linear relationship was found, but the proportionality constant vary from system to system. In particular, Dunlop and Stroink found $A = 14.9$ T/$\mu_B$ for a series of $Fe_{79}SnB_{20-x}Si_x$ with $0 \leq x \leq 12$ [46]. This figure slightly contrasts with the value of 13 T/$\mu_B$ found for similar alloys by Kemeny et al [3]. On the other hand $A = 14$ T/$\mu_B$ was deduced from the results measured on $Fe_xCr_{100-x}B_{20}$ alloys with $50 \leq x \leq 100$ [4]. Panissod et al. found $A = 12.5$ T/$\mu_B$ after compiling the data on a number of amorphous alloys [5]. It seems that for many amorphous alloys the magnetic hyperfine field is in a good approximation proportional to the Fe site magnetic moment and the proportionality constant lies within a relatively narrow range of 12.5 – 15 T/$\mu_B$.

### 5. Conclusions

Based on the results presented in this paper, the following conclusions can be drawn:

1. The magnetic hyperfine field is, in general, not lineraly correlated with the underlying magnetic moment, and the actual correlation depends on the alloy or compound system.
2. Consequently, there is no one unique value of the scaling constant between the two quantities.
3. Instead, it is characteristic of a given alloy or compound system, and for a given system it depends on its composition.
4. For ordered systems the hyperfine coupling constant is characteristic of a crystallographic site, but for a given site it also shows a compositional dependence.